\documentclass[prl,twocolumn,floatfix]{revtex4-1} 

\usepackage{amsmath}  
\usepackage{amsfonts} 
\usepackage{graphicx} 
\usepackage{longtable}  

\newcommand{\Punit}{\ensuremath{{\mbox m}^3\,\mbox{kg}^{-1}\,\mbox{s}^{-2}}}

\newcommand{\Gunit}{\ensuremath{\times 10^{-11}\,\Punit}}

\begin{document}

\title{Recent measurements of  the gravitational constant as a function of time}

\author{S. Schlamminger}
\affiliation{Physical Measurement Laboratory, National Institute of Standards and Technology, Gaithersburg, MD 20899, USA}

\author{J.H. Gundlach}
\affiliation{Center for Experimental Physics and Astrophysics, University of Washington, Seattle, WA 98195, USA}

\author{R.D. Newman}
\affiliation{Department of Physics, University of California Irvine, Irvine, CA 92697-4575, USA}

\date{\today}

\begin{abstract}
A recent publication (J.D. Anderson {\it et. al.}, {\it EPL} {\bf 110}, 1002) presented a strong correlation between the measured values of the gravitational constant $G$ and the 5.9 year oscillation of the length of day.  Here, we compile published measurements of $G$  of the last 35 years.  A least squares regression to a sinusoid with period 5.9 years still yields a better fit than a straight line.  However, our additions and corrections to the G data reported by Anderson {\it et al.} significantly weaken the correlation.

\end{abstract}

\maketitle

\section{Introduction} 

A recent article~\cite{Anderson15} suggests a correlation between measurements of the gravitational constant, $G$,  and the length of day. Figure~1 in~\cite{Anderson15} shows 13 measurements of $G$ as a function of time. Superimposed is a sinusoidal fit with an offset of $\bar{G}=6.673\,90\,\Gunit$, a period $T=5.9$\,years and amplitude $A=0.0016\,\Gunit$. The ratio of amplitude to offset is $2.43\times 10^{-4}$. A second trace shows a scaled version of the change in the length of day, almost indistinguishable from the fit, suggesting a strong correlation of $G$ measurements around the world and the observed change in the length of day.

\begin{figure*}[t!]
\centering
\includegraphics[angle=-90,width=6in]{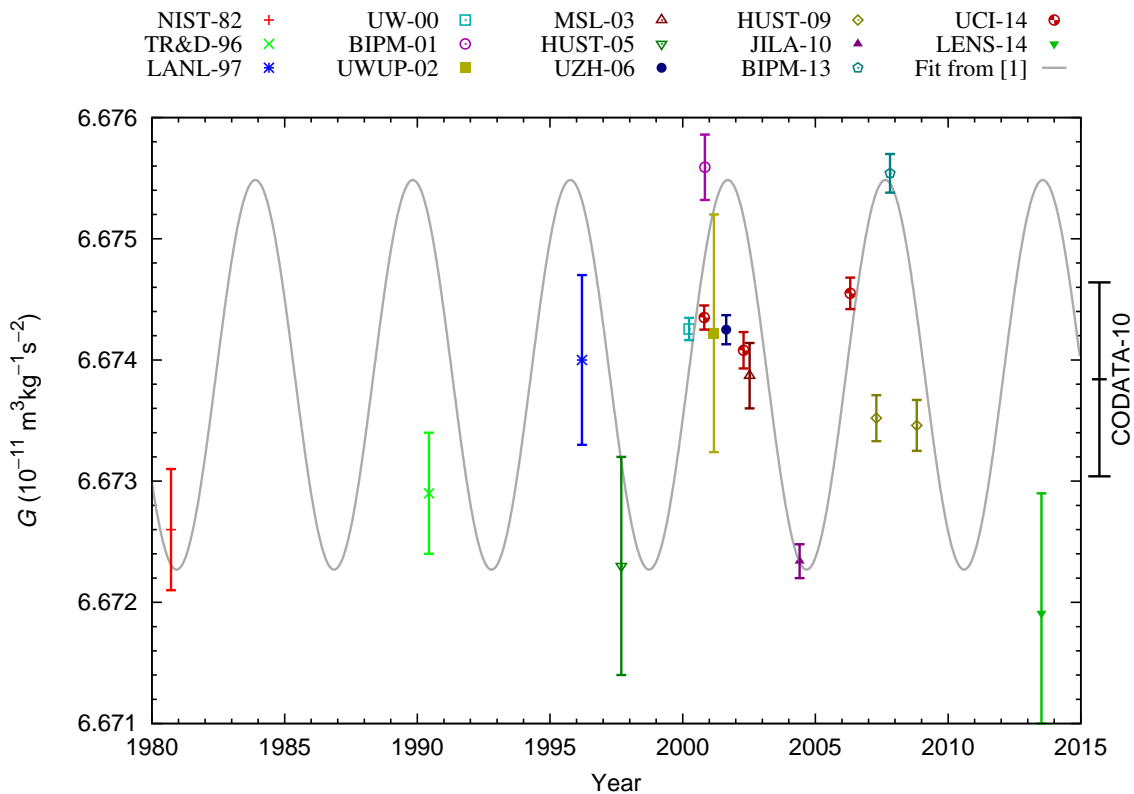}
\caption{Measurements of the gravitational constant, $G$, as a function of time. The TR\&D-96 data were taken over ten years; for this plot the final TR\&D-96 result is shown at the average of their measurement dates. The solid gray  sinusoidal curve is the fit to the data as it appears in \cite{Anderson15}; it is indistinguishable from the scaled length-of-day-variation in the same reference. The point outside the frame gives the 2010 recommended value of $G$ with 1-sigma uncertainties according to the Task Group on Fundamental Constants of CODATA~\cite{CODATA10}.} 
\label{fig1}
\end{figure*}

However, several points in [1] are not plotted at the right time and one experiment ~\cite{Newman14}  is missing.  Here, we provide updated measured values of $G$ with their measurement dates, as displayed in Fig 1. 
\section{Data sources}
It is sometimes difficult to determine the exact time of data acquisition of a published $G$ measurement. Below we attempt to assign a best weighted average of the measurement times involved in each of the most precise $G$ measurements in the last 35 years.
In some  cases, this date is the mean of start and end date of the data acquisition period, in others, it is an average of individual dates when data was taken.  This may not always be the best measure of the effective measurement time; in fitting data we suggest assigning an uncertainty for each tabulated time equal to 20\,\% of the time span.

{\bf NIST-82}:  This experiment was performed at the National Institute of Standards and Technology (then the  National Bureau of Standards) in Gaithersburg, Maryland. A torsion balance used the so-called time-of-swing method in which torsional period is measured in at least two source mass configurations. $G$  is calculated from the difference in the squares of the periods and known mass distributions. The resulting $G=(6.672\,6 \pm 0.000\,5)\Gunit$, was published in 1982~\cite{Luther82}. The measurement dates can be inferred from Table~1 in \cite{Luther80}. The first measurement was August 29 and the last October 10 1980.  We use the average value, September 19 1980, as the time coordinate for this measurement.

{\bf TR\&D-96}: This measurement, performed in Moscow by researchers at Tribotech Research and Development Company, also used a torsion balance in the time-of-swing mode, yielding $G=(6.672\,9 \pm 0.000\,5)\Gunit$, published in \cite{Karagioz96}. The results of measurements spanning 10 years are given  in Table 3 of~\cite{Karagioz96}. Unfortunately the data is given to only four decimal places.  We reproduce the raw data with type A uncertainties in Table~\ref{tab1}. 

\begin{figure}
\centering
\includegraphics[angle=-90,width=3in]{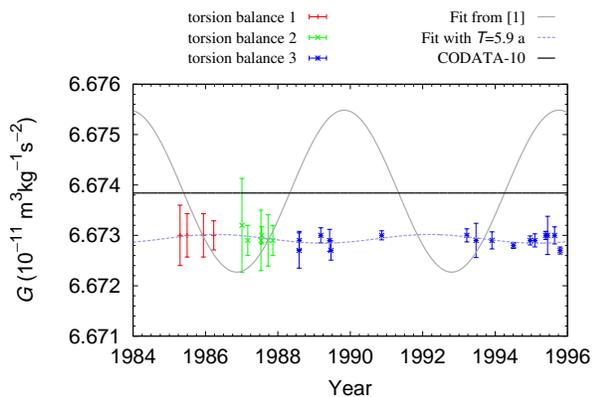}
\caption{Data from~\cite{Karagioz96}. Karagioz and Izmailov measured over a decade using three different torsion balances. Plotted uncertainties are type-A only. According to Ref.~\cite{Karagioz96} the type B uncertainty associated with this experiment is $0.000\,52\Gunit$. }
\label{fig2}
\end{figure}

The TR\&D-96 data alone permits a powerful test for a dependence of $G$ on length of day. Figure~\ref{fig2} shows the data (again with only type A uncertainties) as a function of time. The best fit to a sinusoid with period 5.9 years yields an offset $\bar{G}=6.672\,93\,\Gunit$ and amplitude of $A=0.000\,086\,\Gunit$ with uncertainty $\sigma_A=0.000\,055\,\Gunit$. There are 23 degrees of freedom and the $\chi^2$ is 14.3. Compared to the fit to a full $G$ data set in~\cite{Anderson15}, this fit yields an amplitude smaller by a factor of 19 and phase differing by about 125 degrees.

In 2009, analysis of various correlations of the TR\&D measurements to solar activity and other cosmic periods was published~\cite{Parkhomov09}. Correlations were found, but  were attributed to terrestrial effects --- most probably variations in temperature and the microseismic environment. In~\cite{Parkhomov09} data are shown ranging from 1985 to 2003. Unfortunately the data from 1995 to 2003 is not available to us.

\begin{table}
\begin{tabular}{rlll}
\multicolumn{1}{c}{Date} &
&\multicolumn{1}{c}{$G\times 10^{11}$}
&\multicolumn{1}{c}{$\sigma_G\times 10^{11}$}\\

mm/dd/yyyy&
&\multicolumn{1}{c}{$\left( \Punit \right)$ } 
&\multicolumn{1}{c}{$\left( \Punit \right)$ } 
 \\
\hline
04/19/1985	&\rule{1em}{0pt}&	6.673\,0	&	0.000\,60		\\
06/29/1985	&&	6.673\,0	&	0.000\,43		\\
12/11/1985	&&	6.673\,0	&	0.000\,43		\\
03/25/1986	&&	6.673\,0	&	0.000\,29		\\
01/04/1987	&&	6.673\,2	&	0.000\,93		\\
03/03/1987	&&	6.672\,9	&	0.000\,30		\\
07/14/1987	&&	6.672\,9	&	0.000\,60		\\
07/22/1987	&&	6.673\,0	&	0.000\,17		\\
09/24/1987	&&	6.672\,9	&	0.000\,51		\\
11/11/1987	&&	6.672\,9	&	0.000\,30		\\
08/02/1988	&&	6.672\,7	&	0.000\,35		\\
08/05/1988	&&	6.672\,9	&	0.000\,18		\\
03/09/1989	&&	6.673\,0	&	0.000\,15		\\
06/06/1989	&&	6.672\,9	&	0.000\,22		\\
06/20/1989	&&	6.672\,7	&	0.000\,19		\\
11/13/1990	&&	6.673\,0	&	0.000\,09		\\
03/21/1993	&&	6.673\,0	&	0.000\,13		\\
06/22/1993	&&	6.672\,9	&	0.000\,34		\\
11/30/1993	&&	6.672\,9	&	0.000\,17		\\
07/05/1994	&&	6.672\,8	&	0.000\,06		\\
12/20/1994	&&	6.672\,9	&	0.000\,09		\\
02/06/1995	&&	6.672\,9	&	0.000\,13		\\
05/25/1995	&&	6.673\,0	&	0.000\,08		\\
06/14/1995	&&	6.673\,0	&	0.000\,38		\\
08/24/1995	&&	6.673\,0	&	0.000\,17		\\
10/19/1995	&&	6.672\,7	&	0.000\,07		\\
\end{tabular}
\caption{Measurements
and type A uncertainties from Table~3 of Ref.~\cite{Karagioz96}, with uncertainties converted to absolute values.} 
\label{tab1}
\end{table}

The TR\&D-96 data can be averaged to yield a single data point as displayed in Fig.~\ref{fig1}. The average of the dates listed in table~\ref{tab1} is June 9th 1990.

{\bf LANL-97}: A time-of-swing experiment was performed at the Los Alamos National Laboratory in Los Alamos, New Mexico, yielding $G=(6.674\,0 \pm 0.000\,7)\Gunit$ ~\cite{Bagley97}. The article gives no indication of when the data were taken.  The thesis of C.H. Bagley~\cite{Bagley97a} gives some information. Written on page 15 is ``In January of 1996, I attempted a trial Heyl-type determination with this arrangement, hoping for a percent number or better''. Later it is described how this measurement was much more precise, yielding the final value. On page 71 the reader learns that certain disturbances in the experiment became more frequent as the ambient temperature rose in April and May, until the data became unusable. The thesis was signed July 8 1996. Thus we take March 15 1996 as a time stamp for this data point.

{\bf UW-00}: The measurement with the smallest uncertainty to date was performed at the University of Washington in Seattle, Washington, published in 2000~\cite{Gundlach00}. The rotation rate of a turntable supporting a torsion balance was varied such that the torsion fiber did not twist. In this angular-acceleration-feedback-mode the gravitational acceleration of a torsion pendulum towards source masses is fed back to the turntable, leaving the torsion balance motionless with respect to the turntable and adding the gravitational acceleration to the turntable motion. The gravitational constant is inferred from the second  time derivative of the angle readout of the turntable with respect to time. The value published in 2000 must be slightly corrected due to an originally unconsidered effect of a small mass at the top of the torsion fiber which was also subject to the angular acceleration. This correction is described in~\cite{CODATA02}. After correction, the final result is $G=(6.674\,255 \pm 0.000\,092)\Gunit$.  The times are documented in~\cite{CODATA02}. Two sets of data were taken, one March 10 2000 to  April 1 2000, the other April 3 2000 to April 18 2000. We use March 29 2000 to locate this $G$ value.

{\bf BIPM-01}: These measurements used the first torsion pendulum built at the Bureau International des Poids et Mesures (BIPM) located in S\`{e}vres, near Paris. The experiment measured $G$ with the same instrument operating with two methods. In the Cavendish method, the excursion of a torsion pendulum is measured for two source mass positions.
The corresponding torques are obtained using a torsion constant determined from  the balance's angular moment of inertia and free angular frequency.  

 In the electrostatic servo method, gravitational torque on the pendulum is compensated by an electrostatic torque produced by an electric potential applied to a capacitor with one plate on the pendulum bob and the other fixed. In this phase, the applied voltage is measured.  A calibration experiment measured the capacitance as a function of pendulum angle. 
Combining the results of both methods yielded $G=(6.675\,59 \pm 0.000\,27)\Gunit$ ~\cite{Quinn01}. The results of the Cavendish mode and servo mode are in close agreement, with $G=(6.675\,65 \pm 0.000\,45)\Gunit$ from  the Cavendish mode and $G=(6.675\,53 \pm 0.000\,40)\Gunit$ from the servo mode. According to the authors~\cite{Quinn15}, the servo data were obtained from September 29 to November 2 2000 and the Cavendish data from November 25 to December 13  2000. We take the effective date for the combined $G$ to be the average of the above dates.

{\bf UWUP-02}: This experiment was located at the University of Wuppertal in Germany. The separation of two simple pendulums was measured with microwave interferometry. The forces on the pendulums and, hence, their separation was modulated by external moving source masses. The final value of this measurement, $G=(6.674\,22 \pm 0.000\,98)\Gunit$ is  published in a PhD thesis~\cite{Kleinevoss02}. The appendix  lists the data sets used for the final value. The first data set started January 12 2001, the last ended June 29 2001.  Twelve data sets ranging in duration from 1 to 6 days were taken, mostly within a week of each other. A longer break occurred between March 7 and May 11 and between May 18 and June 25. Averaging the dates of the sets yields March 6.

{\bf MSL-03}: This measurement, performed at the Measurement Standards Laboratory (MSL) of New Zealand, is the only recent measurement performed in the southern hemisphere. It employs a torsion balance in electrostatic servo mode with one difference: The calibration of the capacitance gradient is performed in an angular-acceleration experiment. The final value is $G=(6.673\,87 \pm 0.000\,27)\Gunit$ ~\cite{Armstrong03}. One author~\cite{Armstrong15} informed us that the data was gathered between March 21 2002 and November 1 2002. The average of these dates is  July 11 2002.

{\bf HUST-05}: This is the first measurement of $G$ performed at the Huazhong University of Science and Technology in Wuhan, China. A torsion balance in time-of-swing mode was used. A $G$ value published in 1999~\cite{Luo99}  was subsequently corrected in 2005 for two small errors in mass distribution, yielding $G=(6.672\,3\pm 0.000\,9)\Gunit$.  Data dates without years are given in the 1999 publication: seven sets of measurements were taken, the first starting on August 4 and the last ending on October 15.  The authors report ~\cite{Lu15} that the year was 1997.

We associate September 9 1997, equidistant in time from the start and end of the sets, with the HUST-05 $G$ measurement.

{\bf UZH-06}: The experiment, performed by researchers at the University of Z\"urich, was located at the Paul Scherrer Institute near Villigen Switzerland. The gravitational force of a large mercury mass on two copper cylinders was measured with a modified commercial mass comparator, yielding $G=(6.674\,252 \pm 0.000\,12)\Gunit$, published in 2006~\cite{Schlamminger06}. Figure~8 in this publication shows 43 days of data beginning July 31 2001 and ending September 9 and including a 6 day break.  We take  August 21 2001, as the effective date of this $G$ measurement.

{\bf HUST-09}: A second torsion pendulum apparatus was constructed at HUST and used in time-of-swing mode to make two separate $G$ measurements, whose averaged value $G=(6.673\,49 \pm 0.000\,18)\Gunit$ was first published in 2009~\cite{Luo09}. A long article on the same measurements was published in 2010~\cite{Tu10}, including  the dates of the data sets used in the two experiments. The first experiment consisted of ten sets taken between March  21 2007 and May 20 2007. The second experiment started on October 8 2008 and ended on November 16 2008. The results for the first and second experiments are $G=(6.673\,52 \pm 0.000\,19)\Gunit$ and $G=(6.673\,46 \pm 0.000\,21)\Gunit$, respectively. Averaging the start and end dates of the sets, we obtain April 20 2007 and October 27 2008, respectively.

{\bf JILA-10}: This experiment was performed at the Joint Institute for Laboratory Astrophysics in Boulder, Colorado. Similar to UWUP-02, two simple pendulums with separation determined by a laser interferometer were used to measure $G$, yielding $G=(6.672\,34 \pm 0.000\,14)\Gunit$, reported in 2010 ~\cite{Parks10}. Figure~2 in this report and a table in ~\cite{Parks14} show obtained values of $G$ as a function of time.   Thirteen $G$ values were obtained in a time range May 12 to June 6 2004. Averaging the 13 dates yields May 28 2004.

\begin{table*}[ht!]
\begin{tabular}{llllllrll}
Identifier 
&\multicolumn{1}{c}{$G\times 10^{11}$}
&\multicolumn{1}{c}{$\sigma_G\times 10^{11}$}
&\multicolumn{3}{c}{Data acquisition}&\multicolumn{1}{c}{$e-s$}
& \multicolumn{1}{c}{Device}
& \multicolumn{1}{c}{Mode} 
\\
&\multicolumn{1}{c}{$\left( \Punit \right)$ } 
&\multicolumn{1}{c}{$\left( \Punit \right)$ } 
& Start & End & Average& (Days)
& \\
\hline
NIST-82 & \rule{2em}{0pt}$6.672\,6$ & \rule{1em}{0pt}$0.000\,5$ 
& 08/29/1980 & 10/10/1980 & 09/19/1980 & 42\rule{1em}{0pt}&
 torsion balance&time-of-swing\\
TR\&D-96 & \rule{2em}{0pt}$6.672\,9$ & \rule{1em}{0pt}$0.000\,5$ 
& 04/19/1985 & 10/19/1995& 06/09/1990& 3835\rule{1em}{0pt}&
 torsion balance&time-of-swing \\
LANL-97 & \rule{2em}{0pt}$6.674\,0$ &  \rule{1em}{0pt}$0.000\,7$ 
& 01/01/1996 & 05/31/1996 &03/15/1996 & 151\rule{1em}{0pt}& 
torsion balance& time-of-swing\\
UW-00 &  \rule{2em}{0pt}$6.674\,255$ &  \rule{1em}{0pt}$0.000\,092$ 
& 03/10/2000 & 04/18/2000&03/29/2000& 39\rule{1em}{0pt}&
 torsion balance & acceleration servo\\
BIPM-01s &  \rule{2em}{0pt}$6.675\,53$ &  \rule{1em}{0pt}$0.000\,40$ 
& 09/29/2000 & 11/02/2000&10/16/2000 & 34\rule{1em}{0pt}&
  torsion balance &electrostatic servo\\
BIPM-01c &  \rule{2em}{0pt}$6.675\,65$ &  \rule{1em}{0pt}$0.000\,45$ 
& 11/25/2000 & 12/13/2000& 12/04/2000& 18\rule{1em}{0pt}&
 torsion balance &Cavendish\\
BIPM-01sc &  \rule{2em}{0pt}$6.675\,59$ &  \rule{1em}{0pt}$0.000\,27$ 
& 09/29/2000 & 12/13/2000& 11/02/2000& 75\rule{1em}{0pt}&
 torsion balance &Cavendish \& servo\\
UWUP-02 &  \rule{2em}{0pt}$6.674\,22$ &  \rule{1em}{0pt}$0.000\,98$ 
& 01/12/2001 & 06/29/2001& 03/06/2001& 168\rule{1em}{0pt}&
\multicolumn{1}{l}{two pendulums}\\
MSL-03 &  \rule{2em}{0pt}$6.673\,87$ &  \rule{1em}{0pt}$0.000\,27$ 
& 03/21/2002 & 11/01/2002&07/11/2002 & 225\rule{1em}{0pt}&
 torsion balance &electrostatic servo\\
HUST-05 &  \rule{2em}{0pt}$6.672\,3$ &  \rule{1em}{0pt}$0.000\,9$ 
& 08/04/1997 & 10/15/1997&09/09/1997 & 72\rule{1em}{0pt}&
 torsion balance& time-of-swing\\
UZH-06 &  \rule{2em}{0pt}$6.674\,25$ &  \rule{1em}{0pt}$0.000\,12$ 
& 07/31/2001 & 08/21/2001& 08/21/2001 &21\rule{1em}{0pt}&
\multicolumn{1}{l}{beam balance} \\
HUST-09a &  \rule{2em}{0pt}$6.673\,52$ &  \rule{1em}{0pt}$0.000\,19$ 
& 03/21/2007 & 05/20/2007&04/20/2007  & 60\rule{1em}{0pt}&
 torsion balance& time-of-swing \\
HUST-09b &  \rule{2em}{0pt}$6.673\,46$ &  \rule{1em}{0pt}$0.000\,21$ 
& 10/08/2008 & 11/16/2008&10/27/2008 & 39\rule{1em}{0pt}&
 torsion balance& time-of-swing \\
JILA-10 &  $\rule{2em}{0pt}6.672\,34$ &  \rule{1em}{0pt}$0.000\,14$ 
& 05/12/2004 & 06/06/2004& 05/28/2004 &25\rule{1em}{0pt}&
\multicolumn{1}{l}{two pendulums}\\
BIPM-13s &  $\rule{2em}{0pt}6.675\,15$ &  \rule{1em}{0pt}$0.000\,41$ 
& 11/08/2007 & 01/16/2008&12/15/2007  &69\rule{1em}{0pt}&
 torsion balance &electrostatic servo\\
BIPM-13c &  $\rule{2em}{0pt}6.675\,86$ &  \rule{1em}{0pt}$0.000\,36$ 
& 08/31/2007 & 09/10/2007&09/05/2007 &10\rule{1em}{0pt}&
 torsion balance &Cavendish\\
 BIPM-13sc &  $\rule{2em}{0pt}6.675\,54$ &  \rule{1em}{0pt}$0.000\,16$ 
& 08/31/2007 & 01/16/2008&10/25/2007  &138\rule{1em}{0pt}&
 torsion balance &Cavendish \& servo\\
UCI-14a &  $\rule{2em}{0pt}6.674\,35$ &  \rule{1em}{0pt}$0.000\,10$ 
& 10/04/2000 & 11/11/2000& 10/23/2000&38\rule{1em}{0pt}&
 torsion balance& time-of-swing\\
UCI-14b &  $\rule{2em}{0pt}6.674\,08$ &  \rule{1em}{0pt}$0.000\,15$ 
& 03/25/2002 & 05/12/2002&04/18/2002 &48\rule{1em}{0pt}&
 torsion balance& time-of-swing\\
UCI-14c &  $\rule{2em}{0pt}6.674\,55$ &  \rule{1em}{0pt}$0.000\,13$ 
& 04/08/2006 & 05/14/2006&04/26/2006 &36\rule{1em}{0pt}&
 torsion balance& time-of-swing\\
LENS-14 &  $\rule{2em}{0pt}6.671\,91$ &  \rule{1em}{0pt}$0.000\,99$ 
& 07/05/2013 & 07/12/2013&07/08/2013 &7\rule{1em}{0pt}&
 \multicolumn{1}{l}{atom interferometer}\\
\end{tabular}
\caption{Summary of the most precise measurements of $G$ carried out in the last 35 years. The ``Start'' and ``End'' columns indicate our best estimate of the dates when data acquisition began and ended. The ``Average'' column shows our best estimate for the mean date of data acquisition.  The ``$e-s$'' column gives the difference in days between end and start of data acquisition, important in estimating the amount by which a short-period signal is attenuated. We suggest 20\,\% of the $e-s$ duration number as a meaningful estimate of date  uncertainty.   We separate the two BIPM measurements into four measurements to emphasize that two different methods were used, and include data labeled BIPM-01sc and BIPM-13sc for the best $G$ and dates combining the two methods. Particularly for the 2013 BIPM data, results with the separate methods had strongly anti-correlated uncertainties, so that a $G(t)$ fit using the combined $G$ value can give a significantly different result from a fit treating results from the two methods separately.  The  BIPM data points in Figure 1 represent the combined $G$ data BIPM-01sc and BIPM-13sc.}
\label{tab2}
\end{table*}

{\bf BIPM-13}: At the BIPM, a second torsion balance was constructed to measure $G$ with two different methods. Results were published in 2013~\cite{Quinn13}. Combining the results of both methods yielded $G=(6.675\,54 \pm 0.000\,16)\Gunit$. The Cavendish and servo methods yielded $G=(6.675\,86 \pm 0.000\,36)\Gunit$ and $G=(6.675\,15 \pm 0.000\,41)\Gunit$, respectively. These numbers include a small correction published in an erratum in 2014~\cite{Quinn13}. Per one of the authors~\cite{Quinn15}, the Cavendish data were obtained from August 31 to September 10 2007 and the servo mode data were measured in two campaigns, with November 8, 13, 14, and 16 in 2007 for the first campaign and January 11, 12, 13, 15, and 16 in 2008 for the second campaign. Averaging these dates we obtain October 25 2007 as an effective time stamp for the BIPM-13 data.

{\bf UCI-14}: These measurements, performed using a torsion balance at cryogenic temperatures in time-of-swing mode,  were made near Hanford, Washington. Three types of fibers with differing mechanical properties, especially amplitude dependence of the mechanical losses,  were used. A result for each fiber was published in 2014~\cite{Newman14}: $G=(6.674\,35 \pm 0.000\,10)\Gunit$, $G=(6.674\,08 \pm 0.000\,15)\Gunit$, and $G=(6.674\,55 \pm 0.000\,13)\Gunit$. The principal investigator provided the following time information:  Data with the first fiber was first were obtained from October 4 2000 to November 11 2000. The average of these dates is October 23 2000. Data with the second fiber were obtained during two disjoint intervals. About 14\,\% of the data were obtained between December 8 and December 14 2000, The remainder between March 25 and May 12 2002. For simplicity we assign the average of the dates in 2002, i.e, April 18 2002 to the result with the second fiber. The true average of all dates for this fiber would be roughly January 30 2002. Measurements with the third fiber were collected from April 8 to May 15 2006. The mean of this interval is April 26 2006.

{\bf LENS-14}: Following pioneering work at Stanford University~\cite{Kasevich91}, a precision measurement of $G$ using  a vertical atom interferometer was performed at the University of Florence, Italy. The phase shift between two paths is measured with two source mass configurations.  The $G$ determined from the known mass distributions and the difference of the two phase, is  $G=(6.671\,91 \pm 0.000\,99)\Gunit$,  published in 2014~\cite{Rosi14}.  A longer account of the experiment  appears in~\cite{Prevedelli14},   which states that data was taken between July 5 and July 12 2013. The average of start and end date is July 8 2013. The experiment is on-going  targeting an improved measurement of $G$.

In Table~\ref{tab2} we summarize the precision measurements of big G in the last 35 years.

\subsection{Discussion}

The main purpose of this article is to provide an as complete as possible list of $G$ values determined since 1980, while attempting to assign an as accurate as possible effective date for each measurement, providing data for further investigations similar to that of Anderson and collaborators.

\begin{table*}[ht!]
\begin{tabular}{lrrrrrrlrl}
Fit function &\multicolumn{1}{c}{$T$}
& \multicolumn{2}{c} {\rule{1em}{0pt}$A \times 10^{15}$\rule{1em}{0pt}}
& \rule{1em}{0pt}$\bar{G} \times 10^{11}$\rule{1em}{0pt}
& \multicolumn{1}{c} {Maximum}
&  \multicolumn{1}{c} {$\chi_f^2$} & NDF & $P(\chi^2\ge\chi^2_f)$ & \multicolumn{1}{c}{Remarks}\\
& (years) &
\multicolumn{2}{c}{$\left( \Punit \right)$ } &
\multicolumn{1}{c}{$\left( \Punit \right)$ } &  \\
\hline
from Fig.~1 in [1]  & 5.93 & \rule{3em}{0pt} 16.1& \rule{1em}{0pt}  & 6.673\,88\rule{1em}{0pt}&
 09/13/01\rule{1em}{0pt}& 381\rule{1em}{0pt}  & 14 &$10^{-72}\rule{1em}{0pt}$\\
sine, fixed $T$     & 5.93 & 10.7& \rule{1em}{0pt}  & 6.673\,59\rule{1em}{0pt}&
 03/14/01\rule{1em}{0pt}& 132\rule{1em}{0pt}  & 14 &$10^{-21}\rule{1em}{0pt}$\\
sine, $T$ free      & 0.77 & 11.2& \rule{1em}{0pt}  & 6.673\,58\rule{1em}{0pt}&
 02/21/00\rule{1em}{0pt}& 77\rule{1em}{0pt}  & 13 &$10^{-11}\rule{1em}{0pt}$ & global $\chi^2$ minimum\\
sine, $T$ free      & 6.17 & 11.0& \rule{1em}{0pt}  & 6.673\,54\rule{1em}{0pt}&
 02/13/01\rule{1em}{0pt}& 124\rule{1em}{0pt}  & 13 &$10^{-19}\rule{1em}{0pt}$ & local $\chi^2$ minimum\\
straight line       & n.a. & n.a.& \rule{1em}{0pt}  & 6.674\,13\rule{1em}{0pt}&
 n.a.\rule{1em}{0pt}  & 335\rule{1em}{0pt}  & 16 &$10^{-61}\rule{1em}{0pt}$\\
\end{tabular}
\caption{Fits to  the $G$ data. Here the L2-norm is used exclusively. The ``Maximum'' column  gives the date of the first maximum after 01/01/2000. The ``NDF'' column shows  fits' degrees of freedom.}
\label{tab3}
\end{table*}

We caution users of these data that it is very possible that much or all of the apparent $G$ time variation simply reflects overlooked systematic error, with underestimated systematic uncertainty.

However, we have ventured to make the following fits to data presented in this article, using  the combined numbers for the two BIPM experiments. 
\begin{enumerate}
\item  A sinusoidal function with the parameters found in reference~[1].
\item  A sinusoidal function with free amplitude and phase but period fixed at 5.9 years.
\item  A sinusoidal function with free amplitude, phase and period.
\item  A single time-independent parameter, $\bar{G}$.
\end{enumerate}

Results of these fits are presented in Table~\ref{tab3}. These fits ignored uncertainties in date. Including uncertainties in both coordinates did not significantly affect fit results.

\begin{figure}
\centering
\includegraphics[angle=-90,width=3in]{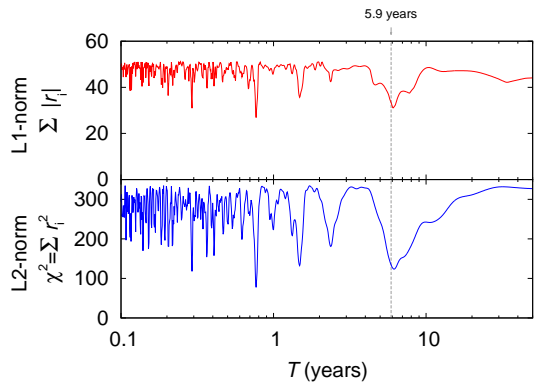}
\caption{ Goodness of fit as a function of period $T$. The upper graph shows the sum of the absolute residual: $\sum_i |r_i|$. The  lower graph shows the sum of the squared residual: $\sum_i r_i^2$. The individual residual is given in both cases by $r_i=\left( G_i - \bar{G} - C\cos{( 2\pi t_i/T)} - S\sin{(2\pi t_i/T)}\right)/\sigma_i$, where $G_i$ and $\sigma_i$ are determined at time $t_i$.}
\label{fig3}
\end{figure}

Figure~\ref{fig3} displays goodness of fit using two different norms for sinusoidal fits as $T$ is varied. The upper and lower graph show fits obtained by minimizing the sum of the absolute residual (L1-norm), $\sum_i |r_i|$, and  the sum of the squared residual (L2-norm), $\sum_i r_i^2$, respectively. Here, $r_i$ is the residual of the i'th data point, given by $r_i=\left( G_i - \bar{G} - C\cos{(2\pi t_i /T)} - S\sin{(2\pi t_i /T)}\right)/\sigma_i$, where $G_i$ and $\sigma_i$ is the measurement and its uncertainty performed at time $t_i$. Fits using the L1-norm are less sensitive to outliers~\cite{Barrodale68}. 
Of note in this plot are:
\begin{enumerate}
\item	There are a number of local minima.
\item	The lowest L1 and L2-norm are both located at $T=0.769$\, years.
\item	A local minimum is found at 6.1 years and 6.2 years for the L1- and L2-norm, respectively; not far from 5.9 years as found by Anderson {\it et al.}.
\item	There is a tantalizing local minimum in the L2-norm at 0.995 year. â\end{enumerate}

We also made a least squares regression to the data taken over a period of more than ten years by Karagioz and Izmailov~\cite{Karagioz96}, as discussed in the Data Sources section of this paper.

The situation is disturbing --- clearly either some strange influence is affecting most $G$ measurements or, probably more likely, measurements of G since 1980 have unrecognized large systematic errors.  The need for new measurements is clear.

Scientific exchange between groups measuring $G$ is necessary. The new working group on big $G$ under the auspices of  International Union of Pure and Applied Physics (IUPAP) was formed to assist experimenters who are interested in these challenging measurements and wish to discuss and understand each other's experiments.

\subsection{Acknowledgment}
We thank D.B.~Newell and B.N.~Taylor for assistance in locating some of the dates of the big $G$ experiments, and we thank the many G practitioners who provided us their best estimates of their measurement dates.  We particularly thank J.~Anderson and his collaborators for extremely helpful suggestions and data.


\begin{thebibliography}{1}

\bibitem{Anderson15} J.D. Anderson, G. Schubert, V. Trimble, and M.R. Feldman, {\it EPL} {\bf 110}, 1002 (2015).

\bibitem{Newman14} R. Newman, M. Bantel, E. Berg, and W. Cross, {\it Phil. Trans. R. Soc. A} {\bf 372}, 20140025 (2014).

\bibitem{CODATA10} P.J.~Mohr, B.N. Taylor, D.B.~Newell, {\it Rev. Mod. Phys.} {\bf 84}, 1527 (2010).

\bibitem{Luther82} G.G.~Luther and W.R.~Towler, {\it Phys. Rev. Lett.}  {\bf 48}, 121 (1982).

\bibitem{Luther80} G.G.~Luther and W.R.~Towler,   in National Bureau of Standards Special Publication 617, Precision Measurement and Fundamental Constants II (1984) pp. 573.

\bibitem{Karagioz96} O.V.~Karagioz, V.P.~Izmailov, {\it Measurement Techniques} {\bf 39}, 979 (1996).

\bibitem{Parkhomov09}  A.G.~Parkhomov, {\it Gravitation and Cosmology} {\bf 15}, 174 (2009).

\bibitem{Bagley97} C.H. Bagley and G.G.  Luther, {\it Phys. Rev. Lett.} {\bf 78}, 3047 (1997).

\bibitem{Bagley97a} C.H. Bagley ``A Determination of the Newtonian Constant of GravitationUsing the Method of Heyl'' Ph.D. Thesis, University of Colorado, Boulder, Colorado, USA (1996).

\bibitem{Gundlach00} J.H.~Gundlach, S.M.~Merkowitz, {\it Phys. Rev. Lett.} {\bf 85}, 2869 (2000).

\bibitem{CODATA02} P.J.~Mohr and B.N. Taylor,  {\it Rev. Mod. Phys.} {\bf 77}, 1 (2005).

\bibitem{Quinn01} T.J.~Quinn, C.C.~Speake, S.J.~Richman, R.S.~Davis, and A.~Picard, {\it Phys. Rev. Lett.} {\bf 87}, 111101 (2001).

\bibitem{Quinn15} T.J.~Quinn, private communication, 2015.


\bibitem{Kleinevoss02} U.~Kleinevo\ss\   ``Bestimmung der Newtonschen Gravitationskonstanten $G$''. Ph.D. Thesis, University of Wuppertal, Wuppertal, Germany.


\bibitem{Armstrong03} T.R.~Armstrong and M.P.~Fitzgerald, {\it Phys. Rev. Lett.} {\bf 91} 201101 (2003).


\bibitem{Armstrong15} T.R.~Armstrong, private communication, 2015.


\bibitem{Luo99} J.~Luo, Z.-K.~Hu, X.-H.~Fu,S.-H.~Fan, and M.-X.~Tang, {\it Phys. Rev. D} {\bf 59},  042001 (1998).

\bibitem{Lu15} Z.~Lu, private communication, 2015.

\bibitem{Schlamminger06} S. Schlamminger {\it et al.} {\it Phys. Rev. D} {\bf 74}, 082001 (2006).

\bibitem{Hu05} Z.-K. Hu, J.-Q.~Guo, and  J.~Luo, {\it Phys. Rev. D} {\bf 71}, 127505 (2005).

\bibitem{Luo09} J.~Luo, Q.~Liu, L.-C.~Tu, C.-G.~Shao, L.-X.~Liu, S.-Q.~Yang, Q.~Li, and Y.-T.~Zhang, {\it Phys. Rev. Lett.} {\bf 102}, 240801 (2009).

\bibitem{Tu10} L.-C.~Tu, Q.~Li,Q.-L. Wang, C.-G. Shao, S.-Q. Yang, L.-X. Liu, Q.~Liu, and J.~Luo  {\it Phys. Rev. D} {\bf 82}, 022001 (2010).


\bibitem{Parks10} H.V.~Parks and J.E.~Faller,  {\it Phys. Rev. Lett.} {\bf 105}, 110801 (2010).


\bibitem{Parks14} H.V.~Parks and J.E.~Faller, {\it Phil, Trans. R. Soc. A} {\bf 372}, 20140024 (2014).

\bibitem{Quinn13} T. Quinn, H.~Parks, C.~Speake, R.~Davis, {\it Phys. Rev. Lett.} {\bf 111}, 101102 (2013). Erratum. 
{\it Phys. Rev. Lett.} {\bf 113}, 039901(E) (2014).

\bibitem{Newmann14} R. Newman {\it et al.}, {\it Phil, Trans. R. Soc. A} {\bf 372} 20140025 (2014).

\bibitem{Kasevich91} M.~Kasevich and S.~Chu,  {\it Phys. Rev. Lett.} {\bf 67}, 181 (1991). 

\bibitem{Rosi14} G. Rosi  {\it et al.}, {\it Nature} {\bf 510}, 518 (2014). 

\bibitem{Prevedelli14} M.~Prevedelli {\it et al.}, {\it Phil, Trans. R. Soc. A} {\bf 372}, 20140030 (2014).

\bibitem{Barrodale68} I.~Barrodale, {J. Roy. Stat. Soc. Ser. C Appl. Stat.}, {\bf 17} 51 (1968).

\end{thebibliography}
\end{document}